\documentclass[a4paper,11pt]{article}
\usepackage{jheppub} 
\usepackage{physics}
\usepackage{mathtools}
\usepackage{bm}

\preprint{CERN-TH-2026-050}

\title{\boldmath Standard Model tests with smeared experiment and theory}

\author{Andreas Jüttner}
\affiliation{CERN, Theoretical Physics Department, Geneva, Switzerland}
\affiliation{School of Physics and Astronomy, University of Southampton, Southampton SO17 1BJ, UK}

\emailAdd{andreas.juttner@cern.ch}

\abstract{
For Standard Model processes in which on-shell intermediate hadronic states contribute — including inclusive semileptonic decays and long-distance effects in rare exclusive decays such as $D\to \pi \ell\ell$ and $B\to K^{(\ast)}\ell\ell$ — spectral-reconstruction techniques provide a promising route to model-independent lattice QCD predictions for use in phenomenology.
The central ingredient is the computation of the energy-smeared spectral density. Following the continuum and infinite-volume limits, the physical amplitude is recovered as the limit of vanishing smearing width. 
However, achieving sufficiently small smearing for a controlled extrapolation remains a significant challenge for current lattice simulations.
In this paper, we therefore propose Standard-Model tests, in which both experimental results and theory predictions are smeared with finite width, similar to what has previously been done in the literature for experimental and lattice $R$-ratio data in the context of the muon $(g-2)_\mu$.
As concrete examples, we discuss the cases of inclusive meson decay and long-distance contributions to rare semileptonic meson decay. 
}

\begin{document}
\maketitle
\flushbottom
\newpage
\section{Introduction}
\label{sec:Decay rate}
Lattice QCD has over the past decades matured into a precision tool for a broad range of phenomenologically relevant quantities.~\cite{FlavourLatticeAveragingGroupFLAG:2024oxs}. This is particularly so for observables for which a direct relation between the observable in Minkowski and Euclidean space  exists. Examples are the decay constants of stable (in QCD) mesons ($\pi$, $K$, $D_{(s)}$, $B_{(s)}$, \dots), or  the transition matrix-elements of local operators between stable initial and final states ($K\to\pi \ell\bar\nu$, $D\to\pi\ell\bar\nu$, $B\to\pi\ell\bar\nu$, \dots). A more comprehensive listing can be found in the introductory section of the FLAG review~\cite{FlavourLatticeAveragingGroupFLAG:2024oxs}. Together with input from particle-physics experiments  around the world,  lattice-QCD predictions  are playing a  crucial  role in precision tests of the Standard Model (SM). 

Here, we address a class of phenomenologically very interesting  processes and observables for which lattice calculations have not yet reached a similar level of maturity. In particular, we concentrate on cases for which the relation between the observable in Minkowski and Euclidean space is less obvious. One example is rare semileptonic decay, where long-distance effects allow for intermediate-state resonance contributions (e.g. charm resonances in $B\to K^{(\ast)}\ell^+\ell^-$, or $\rho,\,\omega$, and $\phi$ resonances to $D\to \pi \ell^+\ell^-$). A comprehensive analysis of the experimental data eventually requires reliable first-principles predictions for the underlying hadronic physics, rather than theoretical models. Concrete steps towards a lattice computation of such long-distance effects have recently been proposed in~\cite{Frezzotti:2025hif}. Another example is the hadronic tensor entering the SM prediction for inclusive meson decay, which receives contributions from a tower of intermediate states. The decay channel $B_{(s)}\to X_c\ell\bar\nu$ has in this context attracted considerable attention owing to its role in the determination of the CKM matrix element $|V_{cb}|$. Techniques for the lattice computation have been developed rather recently~\cite{Hashimoto:2017wqo,Gambino:2020crt}, with first results for $D_s\to X_{c,s}\ell\bar \nu$ and $B_s\to X_c\ell\bar\nu$ in~\cite{Gambino:2022dvu,Barone:2023tbl,DeSantis:2025yfm,Kellermann:2025pzt}. 

The  difficulty common to these and other processes results from  branch cuts in the hadronic amplitude due to  intermediate hadronic states. They obstruct a naive Wick rotation between  Minkowski and Euclidean space. In Euclidean signature, the presence of the intermediate states leads to contributions to the amplitude that diverge as the infinite-volume limit is taken if conventional lattice methods are used. Although subtracting them is possible in select  cases~\cite{Christ:2015aha,Christ:2016mmq}, it becomes impractical, if not impossible, in the presence of multiple or unstable intermediate states. 
A more promising approach is to reconstruct the spectral density for a given process, in this way avoiding Euclidean-time integration. This idea was first developed in~\cite{Barata:1990rn}, and has gained momentum in the lattice community over the past years in terms of more formal developments~\cite{Hansen:2017mnd,Bruno:2020kyl,DelDebbio:2024lwm,Patella:2024cto,Frezzotti:2025hif} as well as further applications~\cite{Bonanno:2023ljc,Evangelista:2023fmt,Panero:2023zdr,Frezzotti:2023nun,Bennett:2024cqv,Frezzotti:2024kqk}. 
In~\cite{Bulava:2019kbi} it was worked out how energy-smeared spectral densities for scattering amplitudes could be obtained from Euclidean correlation functions at finite smearing width $\epsilon$. The physical amplitude can be obtained in the $\epsilon\to 0$ limit. However, the smearing width $\epsilon$ sets a new infrared scale in the simulation. It was found that for a controlled $\epsilon\to 0$ limit rather large lattice volumes are therefore required, which are computationally prohibitively expensive (\cite{Bulava:2019kbi} estimates $M_\pi L\approx 10-20$, where $L$ is the spatial extent of the lattice). 

Spectral reconstruction for a given process uses the relation between suitably constructed Euclidean correlation functions and the Laplace transform of the corresponding spectral density. The finite volume requires one to consider energy-smeared versions of the discrete spectral density. Since lattice computations only provide information for the correlation function at discrete values of Euclidean time and with finite statistical precision, the reconstruction of the spectral density constitutes an ill-posed inverse problem. The difficulty then lies in reliably controlling systematic effects in the reconstruction. To this end, different methods are being developed and used. The HLT method~\cite{Hansen:2019idp}, like Backus–Gilbert~\cite{Backus:1968svk,Backus:1970}, controls the ill-posed problem by balancing resolution of physical structure (e.g. resonances) in the amplitude against statistical noise, leading to  quantifiable uncertainties. 
In the Chebyshev approach~\cite{Barata:1990rn,Hashimoto:2017wqo,Bailas:2020qmv,Gambino:2020crt}, the spectral function is expanded in Chebyshev polynomials, and the maximum order retained in the expansion sets the resolution: higher order allows for resolving sharper features but amplifies noise. 
Other approaches are being developed, for instance, ones based on Nevanlinna-Pick interpolation~\cite{Bergamaschi:2023xzx,Huang:2023gpb,Fields:2025glg} or the Mellin transform~\cite{Bruno:2024fqc}.

Common to all approaches is the observation that, as opposed to the actual physical amplitude, the smeared amplitude can be computed on the lattice with well-controlled systematic errors. The physical amplitude can then be obtained after taking the continuum and infinite-volume limit, in terms of the extrapolation $\epsilon\to 0$. The requirement of very large volumes for reliably controlling the extrapolation motivates us to ask whether a comparison of experiment and theory could instead be accomplished at finite smearing width, without relying on any model assumptions about the $\epsilon$ extrapolation. This would make lattice predictions for numerous physical processes much more feasible with current resources. The question to address is, are there smearing prescriptions, which applied to experimental data (e.g. differential decay rates) and reconstructed from lattice QCD, respectively, allow for a direct comparison and, most importantly, meaningful SM tests.

This question has been asked before:
In~\cite{ExtendedTwistedMassCollaborationETMC:2022sta} data for the $R$ ratio obtained from experiment for $e^+e^-\to{\rm hadrons}$ and from lattice simulations was smeared with a Gaussian kernel and compared at finite smearing width, identifying a $3\sigma$ tension in the region of the $\rho$-resonance peak. 
In~\cite{Frezzotti:2023nun} it was suggested that experimental data could be interpreted in terms of a theoretical model function, which, in turn, could be subjected to the smearing. The resulting smeared model function could then be compared directly to the smeared lattice data. 

Here, we argue that such a model function is not always required, and that a direct comparison between smeared experimental data and theory predictions is possible at finite smearing width. This turns out to be the case for a number of phenomenologically relevant cases, such as interference of short- and long-distance contributions to rare semileptonic meson decay, or inclusive meson decay. We also discuss cases for which the analytical structure of the hadronic contribution does not permit such a direct comparison. 
For these cases we propose new ideas that allow for meaningful finite-width comparisons between theory and experiment.

Based on the ideas presented here, a joint effort of experimentalists and theorists in analysing smeared data could enable a new and timely analysis of long-distance effects in experimental data in flavour physics and beyond, providing new levers for controlling hitherto hard-to-estimate systematic effects. 

The paper continues in Sec.~\ref{sec:Theoretical background} with a brief theoretical motivation. Sec.~\ref{sec:Decay rate} discusses how a direct comparison of smeared experiment and theory is possible for two generic classes SM processes found in phenomenology. We also discuss situations where the analytical properties of the SM expressions obstructs a direct comparison at finite smearing, and expose the resulting defect  in terms of a resonance model. Sec.~\ref{sec:Some examples} is dedicated to making concrete proposals for SM tests at finite smearing width for the examples of inclusive meson decay and rare semileptonic meson decay, respectively. We present our conclusions in Sec.~\ref{sec:Conclusions}.
\section{Theoretical motivation}
\label{sec:Theoretical background}
We consider processes defined in terms of hadronic matrix elements, in which intermediate states can go on-shell. The corresponding hadronic  amplitude  has a dispersive representation
\begin{align}\label{eq:dispersive amplitude}
    H(x)=&\,\lim\limits_{\epsilon\to 0}\int\limits_{x_{\rm min}}^\infty dx'\,  \frac{\rho(x')}{x'-x-i\epsilon}\,,
\end{align}
where $x$ represents a kinematic variable (e.g. energy or squared momentum transfer), and 
where the spectral density $\rho=\frac 1\pi {\rm Im}H$ (in a distributional sense) is the  discontinuity across the branch cut without support for $x<x_{\rm min}$.  We assume that the amplitude is finite (physical) in terms of a principal-value prescription, which  may require subtractions in Eq.~(\ref{eq:dispersive amplitude}). A well-known example is the spectral density for the process $e^+e^-\to{\rm hadrons}$, and its dispersive relation to the leading hadronic contribution to the lepton anomalous magnetic moment, $(g-2)_\ell$. In this case $x_{\rm min}$ corresponds to the two-pion threshold. 

For a lattice calculation of the {physical} amplitude, zeros in $x'-x$ obstruct the continuation to Euclidean space, and hence, a direct computation of the amplitude. The way out suggested in~\cite{Barata:1990rn,Hansen:2017mnd,Hansen:2019idp,Bruno:2020kyl} is to compute the \emph{harmonic extension} on the lattice, i.e., the amplitude at finite values of  $\epsilon$.
The continuum and finite-volume limits of the energy-smeared amplitude computed in lattice QCD can then be taken, followed by the $\epsilon\to0$ limit, where the physical amplitude is recovered.

A detailed account of how such a computation for the case of electroweak amplitudes might work was given in~\cite{Frezzotti:2023nun}. The paper concludes that two criteria have to be met for a controlled $\epsilon\to 0$ extrapolation. First, the smearing width must cover the separation between the discrete peaks of the finite-volume spectral density. The corresponding constraint is $\epsilon \gg 1/L$, where $L$ is the spatial extent of the lattice volume. Depending on the structure of the spectral density of interest, prohibitively large lattice volumes might be required. 
Second, the smearing width must be smaller than any interval in $x$ over which the spectral density varies significantly, expressed in terms of the logarithmic derivative $\Delta(x)$ of the amplitude $H(x)$ as $\epsilon\ll |\Delta(x)|$. Meeting these conditions in a lattice simulation will be difficult  for the foreseeable future. For instance,  intermediate-state resonance contributions to hadronic transitions (e.g. rare semileptonic meson decay such as $B\to K^{(\ast)} \ell^+\ell^-$ or $D\to \pi \ell^+\ell^-$) pose challenging requirements for a controlled $\epsilon\to 0$ limit.

\section{Smeared decay rate}
\label{sec:Decay rate}
We now consider two generic situations, in which an experimentally 
measurable (real-valued) differential decay rate can be expressed in terms of 
\begin{align}
    {\frac{d\Gamma}{dx}}(x)=\left\{
        \begin{array}{ll}
         \Phi(x)\cdot{\rho}(x)&{\rm (case \, I -\textrm{linear in the spectral density $\rho$})}\,,\\
        H(x)\cdot \Phi(x)\cdot {H}^\ast(x)&{\rm (case \, II -\textrm{quadratic  in the amplitude $H$})}\,,
        \end{array}\right.
\end{align}
where $x$ represents a kinematical variable, and 
where $\Phi$ is a function that can contain normalisations, CKM matrix elements and phases, or kinematic factors, and where  $\rho$, $H$ and $\Phi$ can have Lorentz and/or other indices, allowing for mixing of  short- and long-distance effects.\footnote{Note that $x$ and $\Phi$ likely differ between case I and II.}
An example for case I, as we will later discuss in detail, is the differential decay rate for inclusive meson decay, while an example for case II is rare semileptonic meson decay.
We further assume that a dispersive representation of the amplitude similar to Eq.~(\ref{eq:dispersive amplitude}) exists. For the following discussion it is sufficient to assume $\Phi$, $\rho$ and $H$ to be scalar, i.e., to have no indices.
We  define smeared quantities as
\begin{align}
    \langle f(x)\rangle_\epsilon=\int\limits_{-\infty}^\infty{dx'}\,K_\epsilon(x'-x)f(x')\,\,{\rm with}\,\,\int\limits_{-\infty}^\infty dx' K_\epsilon(x')=1\,,
\end{align} with a suitable smearing kernel $K_\epsilon$ defined shortly.
Smeared versions of above cases I and II, hence, are
\begin{align}
   \left< \frac{d\Gamma}{dx}(x)\right>_\epsilon=\left\{
        \begin{array}{ll}
        \langle \Phi(x)\cdot\rho(x)\rangle_\epsilon&{\rm (case \, I)}\,,\\
        \langle H(x)\cdot \Phi(x)\cdot H^\ast(x)\rangle_\epsilon&{\rm (case \, II)}\,.
        \end{array}\right.\label{eq:classification}
\end{align}
\subsection{Case I -- Smearing observables that are linear in the spectral density}
After smearing in case I with a Poisson kernel
\begin{align}
    K_\epsilon(x'-x)=\frac 1\pi \frac{\epsilon}{(x'-x)^2+\epsilon^2}
    \,,
\end{align}
one finds
\begin{align}\label{eq:Poisson smearing case I}
\langle\Phi(x)  \rho(x) \rangle_\epsilon=&\,\int\limits_{-\infty}^\infty dx'
 K_\epsilon(x'-x)\Phi(x')\rho(x')
 ={\rm Im}\left[\frac 1\pi\int\limits_{x_{\rm min}}^\infty d x'\frac {\Phi(x')\rho(x')}{x'-x-i\epsilon}\right]\,.
\end{align}
The Poisson smearing regulates the physical amplitude in terms of the smearing width $\epsilon$ by \emph{harmonic extension}. In this way  poles along the real axis are avoided for $\epsilon>0$. Since Eq.~(\ref{eq:Poisson smearing case I}) can be obtained from  experimental data for $\Phi(x)  \rho(x)$ as part of the  data analysis, a direct comparison to the same quantity computed in lattice QCD is possible.
In the case of inclusive meson decay (which we will discuss in detail in Sec.~\ref{sec:Inclusive meson decay}), $\Phi$ is a polynomial in $x$, and the effect of smearing can be seen more explicitly -- one finds for $n\in\mathbb{N}$ 
\begin{align}\label{eq:harmonic extension case I}
    \langle x^n\rho(x)\rangle_\epsilon=\,\frac 1\pi {\rm Im}[(x+i\epsilon)^n H(x+i\epsilon)]+\frac 1\pi\sum\limits_{k=0}^{n-1}{\rm Im}[(x+i\epsilon)^k]\,\mu_{n-1-k}\,,
\end{align}
with moments $\mu_k=\int dx\, x^k\rho(x)$. The relation highlights that the smeared differential decay rate depends on the corresponding hadronic amplitude $H$ by means of the dispersion relation Eq.~(\ref{eq:dispersive amplitude}) at finite $\epsilon$, i.e., before the $\epsilon\to 0$ limit. Note that Eq.~(\ref{eq:harmonic extension case I})  also holds in the presence of kinematical constraints (e.g. finite upper integration limit) as encountered in the case of inclusive decay.

For SM processes that are linear in the spectral density (case I) the smeared experimental data and the same quantity reconstructed on the lattice can therefore be compared without the requirement of a $\epsilon\to 0$ limit.
\subsection{Case II -- Smearing observables that are bilinear in the hadronic amplitude}
Let us now have a look at case II, assuming for simplicity $\Phi(x)=1$ and $H(x)$ a scalar hadronic amplitude with dispersive representation Eq.~(\ref{eq:dispersive amplitude}). Then
\begin{align}
    \langle H(x)\rangle_\epsilon=
    \lim_{\eta\to 0}\frac \epsilon\pi \int\limits_{{-\infty}}^\infty \!\!dx''\frac{1}{(x''-x)^2+\epsilon^2}\int\limits_{x_{\rm min}}^\infty \!\!dx'\frac {\rho(x')}{x'-x-i\eta}=\int\limits_{x_{\rm min}}^\infty \!\!dx'\frac {\rho(x')}{x'-x-i\epsilon}=H(x+i\epsilon)\,.
\end{align}
This holds for $H$ analytic in the upper half plane, with $\rho(x)$ decaying sufficiently fast for $x\to\infty$ (otherwise a subtracted dispersion relation can be considered). These conditions are usually fulfilled for physical amplitudes thanks to  causality.
For the norm-squared of Poisson-smeared amplitudes we then obtain
\begin{align}\label{eq:case II a}
    |\langle H(x)\rangle_\epsilon|^2=
    |H(x+i\epsilon)|^2\,.
\end{align}
Experiment however measures the observable $\mathcal{O}\sim HH^\ast$ and smearing of experimental data corresponds to $\langle |H(x)|^2\rangle_\epsilon$, which differs from Eq.~(\ref{eq:case II a}). The expression for $|H(x)|^2$ in terms of the dispersive representation Eq.~(\ref{eq:dispersive amplitude}),
contains the limits of two poles, respectively, above and below the same  point on the real $x$ axis. It does  therefore not satisfy Cauchy-type dispersion relations, and hence, no suitable smearing kernel exists that would yield equality with Eq.~(\ref{eq:case II a}) at $\epsilon >0$. We define the defect term 
\begin{align}\label{eq:defect term}
\Delta_\epsilon(x)=&\,
     \langle |H(x)|^2\rangle _\epsilon
    - |\langle H(x)\rangle_\epsilon|^2
    > 0\qquad{\rm for\,}\epsilon>0\,,
\end{align}
as a measure of the difference.

Fortunately, we do not have to abandon the idea of a model-independent comparison of smeared  experiment and theory altogether:
  observables of case-II structure often receive contributions from both short- and long-distance effects, and physical observables (e.g. CP-odd) can be constructed, in which the pure long-distance term $HH^\ast$ cancels.

Let us therefore now assume a hadronic amplitude with both short- and long-distance contributions
\begin{align}
    H(x)=&\,H_{\rm SD}(x)+H_{\rm LD}(x)\,.
\end{align}
The short-distance contribution originates from local matrix elements and can be expressed in terms of hadronic (e.g. scalar, vector and tensor) form factors. 
After Poisson smearing experimental data we expect
\begin{align}\label{eq:interference term}
    \langle|H(x)|^2\rangle_\epsilon = 
    \langle |H_{\rm SD}(x)|^2\rangle_\epsilon
     + 
    \langle |H_{\rm LD}(x)|^2\rangle_\epsilon+
    2{\rm Re}\langle H^\ast_{\rm SD}(x)H_{\rm LD}(x)\rangle_\epsilon\,.
\end{align}
Regarding the first term on the r.h.s., the form factors entering $H_{\rm SD}$ can be computed in lattice QCD using   standard techniques~\cite{FlavourLatticeAveragingGroupFLAG:2024oxs}. Computing $\langle |H_{\rm SD}(x)|^2\rangle_\epsilon$ should therefore be unproblematic. Compared to resonance contributions, the local form factors vary slowly within the semileptonic region. Depending on the actual size of $\epsilon$, $\langle |H_{\rm SD}(x)|^2\rangle_\epsilon\approx |H_{\rm SD}(x)|^2$ might therefore be sufficient and convenient approximation to be considered. The second term is the pure long-distance contribution, which can be computed on the lattice only up to the defect term $\Delta_\epsilon$ defined in Eq.~(\ref{eq:defect term}).
%
The third term represents interference  of long- and short-distance physics. In it, the amplitude $H_{\rm SD}$ can again be predicted in terms of local form factors, while $H_{\rm LD}$ has to be computed by spectral reconstruction (here it can be worthwhile considering the approximation $\langle H^*_{\rm SD}(x)H_{\rm LD}(x)\rangle_\epsilon\approx H^*_{\rm SD}(x)\langle H_{\rm LD}(x)\rangle_\epsilon$). 
For experimental observables such as  CP asymmetries, where the pure-LD term $\langle |H_{\rm LD}(x)|^2\rangle_\epsilon$ is absent, a fully model-independent comparison of theory and experiment at finite smearing is therefore possible. Since NP effects are expected to be enhanced in interference, the proposal made in this paper opens up interesting new opportunities for testing the SM.
  We then find
  \begin{align}
      \langle {\rm Re}[H^\ast_{\rm SD}(x) H_{\rm LD}(x)]\rangle_\epsilon = \frac \epsilon \pi \int\limits_{-\infty}^\infty dx'\frac{{\rm Re}[H^\ast_{\rm SD}(x') H_{\rm LD}(x')]}{(x'-x)^2+\epsilon^2}=
      {\rm Re}[H^\ast_{\rm SD}(x+i\epsilon)H_{\rm LD}(x+i\epsilon)]\,.
  \end{align}
  This smeared interference term can be determined from smeared experimental data, and it can also be computed in lattice QCD (see example in Sec.~\ref{sec:Semileptonic rare D decay}). 
  Since NP can couple to the SM in interference terms, a model-independent analysis carried out in this way will correctly take into account weak and strong phase effects from first principles,  and would therefore go beyond current analyses based on factorisation.
  
 Interesting information on observables containing the pure long-distance contribution $H_{\rm LD}H_{\rm LD}^\ast$ can however still be gained. In the next section we discuss an analysis making certain  model assumptions for the amplitude $H_{\rm LD}$. Following that we also propose a model-independent analysis involving comparisons of data at different smearing widths, which is meaningful also for the $H_{\rm LD}H_{\rm LD}^\ast$ term.
\subsubsection{Model-dependent SM tests assuming Breit-Wigner }
\label{sec:Assuming BW}
In order to better understand the nature of the defect term $\Delta_\epsilon$ of Eq.~(\ref{eq:defect term}), we now assume a Breit-Wigner ansatz for the hadronic amplitude with coupling $c$, resonance mass $E_0$ and width $\Gamma$,
\begin{align}
    H(E)=\frac{c}{E_0-E-i\Gamma/2}\,.
\end{align}
 For the norm-squared of the Poisson-smeared amplitude  we then find
\begin{align}
    |\langle H(E)\rangle_\epsilon|^2=\frac{|c|^2}{(E_0-E)^2+(\Gamma/2+\epsilon)^2}\,,
\end{align}
exposing that the smearing increases the Breit-Wigner width $\Gamma$.
However, we also find
\begin{align}
    \langle|H(E)|^2\rangle_\epsilon=\frac{|c|^2}{(E_0-E)^2+(\Gamma/2+\epsilon)^2}\left(1+\frac{2\epsilon}{\Gamma}\right)\,,
\end{align}
for the smeared norm-squared, which differs from the previous expression by a correction term that depends on the resonance and smearing widths.
This is the anticipated defect, which can be written compactly as
\begin{align}
    \Delta_\epsilon(E)=| \langle H(E)\rangle_\epsilon|^2\frac {2\epsilon}{\Gamma}\,.
\end{align}
In agreement with intuition, the defect is largest in absolute terms in the vicinity of resonances. Broader resonances lead to smaller effects than narrow ones. 
 Making model assumptions, hence, allows for qualitative as well as quantitative estimates of the defect term.

For the case of multiple interfering resonances contributing to the amplitude we assume a superposition of $N$ Breit-Wigners\footnote{We assume the resonance energies $E_i$ and widths $\Gamma_i$ to be such that negative energies under Poisson smearing are negligible.} 

\begin{align}
    H(E)=\sum\limits_{n=1}^N H_i(E)\,,\;{\rm where}\;\;H_i(E)=\frac{c_i}{E_i-E-i\Gamma_i/2}\,.
\end{align}
We find
\begin{align}
    | \langle H(E)\rangle_\epsilon|^2=&\,\sum\limits_{i=1}^N\frac {|c_i|^2}{(E_i-E)^2+(\Gamma_i/2+\epsilon)^2}
    \nonumber\\
    &\qquad +2\sum\limits_{i<j}{\rm Re}\frac{c_i^\ast c_j}{(E_i-E+i(\Gamma_i/2+\epsilon))
    (E_j-E-i(\Gamma_j/2+\epsilon))}\,,
\end{align}
and
\begin{align}
    \langle |H(E)|^2\rangle _\epsilon=&\,\sum\limits_{i=1}^N\frac {|c_i|^2}{(E_i-E)^2+(\Gamma_i/2+\epsilon)^2}\left(1+\frac {2\epsilon}{\Gamma}\right)
    \nonumber\\
    &+2\sum\limits_{i< j}{\rm Re}\frac{c_i^\ast c_j}{(E_i-E+i(\Gamma_i/2+\epsilon))
    (E_j-E-i(\Gamma_j/2+\epsilon))}\left(1+\frac{2\epsilon}{\Gamma_{ij}-iE_{ij}}\right)\,.
\end{align}
where $\Gamma_{ij}=\Gamma_i+\Gamma_j$ and $E_{ij}=E_i-E_j$.
The defect term
\begin{align}
     \Delta_\epsilon(E)
     =&\,\,\sum\limits_{i=1}^N\frac {|c_i|^2}{(E_i-E)^2+(\Gamma_i/2+\epsilon)^2}\frac {2\epsilon}{\Gamma_i}
    \nonumber\\
    &+\sum\limits_{i\neq j}{\rm Re}\frac{c_i^\ast c_j}{(E_i-E+i(\Gamma_i/2+\epsilon))
    (E_j-E-i(\Gamma_j/2+\epsilon))}\frac{2\epsilon(\Gamma_{ij}+iE_{ij})}{\Gamma_{ij}^2+E_{ij}^2}\,,
\end{align}
 quantifies the fact that correlations between different amplitudes on the scale of the smearing width are not captured in the square of smeared amplitudes. For small smearing width $\epsilon \ll \Gamma_i,\Gamma_k, |E_i-E_j|$ individual resonances are resolved and the defect term is small. For $\epsilon\sim \Gamma_i,\Gamma_j,|E_i-E_j|$ the smearing width overlaps neighbouring resonances, and their interference becomes noticeable. 

We conclude that model assumptions (such as Breit-Wigner or Gounaris-Sakurai~\cite{Gounaris:1968mw}) make a direct comparison of smeared experiment and amplitude in case II  possible and allow for a quantitative as well as qualitative study of the defect term.
\subsubsection{Model-independent SM tests based on the  defect term}
\label{sec:The defect term and testing the SM}
We note that the defect term in Eq.~(\ref{eq:defect term})
is positive  definite for $\epsilon>0$, which is a model-independent statement. 
A negative $\Delta_\epsilon(x)$  for $\langle|H(x)|^2\rangle_\epsilon$ from experiment and $|\langle H(x)\rangle_\epsilon|^2$ from theory would therefore 
indicate some (not-accounted-for) short-distance contribution. We can therefore use the defect term to test the  SM. 
Similarly, 
we can also make use of the fact that Poisson smearing forms a semi group,
\begin{align}
    \langle f\rangle_{\epsilon_1+\epsilon_2}=\langle \langle f\rangle_{\epsilon_1}\rangle_{\epsilon_2}\,.
\end{align}
This provides for another interesting opportunity to test the SM:
Estimate $\Delta_{\epsilon_1}(x)$ as above from experimental and theory data, and predict $\Delta_{\epsilon_1+\epsilon_2}(x)$ by repeated Poisson smearing $\langle \Delta_{\epsilon_1}\rangle_{\epsilon_2}(x)$. Compare this to the estimate of the same quantity but obtained from 
$\langle|H(x)|^2\rangle_{\epsilon_1+\epsilon_2}$ from experiment and $|\langle H(x)\rangle_{\epsilon_1+\epsilon_2}|^2$ from theory. Any disagreement could indicate a problem in the analysis, or, of course, be due to NP.\\

\noindent We conclude that 
\begin{itemize}
\item
for case-II quantities the defect term $\Delta_\epsilon$ constitutes an irreducible obstacle for direct comparisons of smeared observables containing pure-long-distance contributions $\sim H_{\rm LD}H_{\rm LD}^\ast$. One has to rely on model assumptions for the long-distance part (Sec.~\ref{sec:Assuming BW}), or it has to make use of simultaneous analyses at different smearing widths (Sec.~\ref{sec:The defect term and testing the SM}).
\item a direct comparison of experiment and theory at finite smearing is however possible for observables containing only 
pure short-distance and/or short- and long-distance interference contributions at most linear in $H_{\rm LD}$. Model-independent SM tests at finite smearing can therefore be formulated for experimentally accessible observables in which the pure long-distance effects cancel. Examples are CP  or angular asymmetries where  sensitivity to new physics can appear linearly and therefore enhanced in the long-distance contribution. Having a reliable prediction for the long-distance contribution in combination with experimental measurements, hence, allows for reliable constraints on potential  NP scales.
\end{itemize}

\section{Two examples}
\label{sec:Some examples}
In this section we discuss how the ideas presented so far apply to the two examples of inclusive and rare exclusive semileptonic meson decay.  After briefly summarising the SM expressions, we discuss for both cases how a comparison of experiment and theory predictions can be accomplished at finite smearing width.
\subsection{Example I: Inclusive meson decay}
In the inclusive analysis of semileptonic meson decay one chooses an initial-state meson and sums the decay rate over all semileptonic final states allowed by symmetry and kinematics and with a particular choice of flavour content. Until recently, the analysis within the SM was based mainly on the operator product expansion. The comparison of inclusive~\cite{Fael:2018vsp,Fael:2020tow,Belle:2021idw,Belle-II:2022evt,Bernlochner:2022ucr,Finauri:2023kte,Bordone:2021oof} and exclusive~\cite{FermilabLattice:2021cdg,Harrison:2023dzh,Aoki:2023qpa,Martinelli:2021myh,Martinelli:2021onb,Martinelli:2022xir,Martinelli:2023fwm,Bordone:2024weh} analyses is the origin of the $|V_{cb}|$ and $|V_{ub}|$ CKM puzzles -- current determinations of these CKM matrix elements are at tension with each other~\cite{FlavourLatticeAveragingGroupFLAG:2024oxs,ParticleDataGroup:2024cfk,HFLAV:2022pwe}. An independent determination of the inclusive decay rate in lattice QCD~\cite{Hashimoto:2017wqo,Gambino:2020crt,Gambino:2022dvu,Barone:2023tbl,DeSantis:2025yfm,Kellermann:2025pzt}, and hence, an independent determination of CKM-matrix elements, would therefore be highly valuable.
\label{sec:Inclusive meson decay}
\subsubsection{SM expressions}
\label{sec:SM expressions inclusive decay}
We consider inclusive semileptonic $B\to X_c\ell\bar\nu$ decay. The discussion carries over to other inclusive meson-decay channels. The SM differential decay rate  reads
\begin{align*}
 \frac{d^3 \Gamma}{d q^2 d q_0 d E_{l}} = \frac{G^2_F |V_{cb}|^2}{8\pi^3} L^{\mu\nu}(p_l, p_{\nu_l}) W_{\mu\nu}(q) \, .
\end{align*}
The lepton contribution is given in terms of the lepton tensor
\begin{align}
 L^{\mu\nu}(p_l, p_{\nu_l}) = p_{l}^{\mu}p_{\nu_l}^{\nu} +  p_{l}^{\nu}p_{\nu_l}^{\mu} - g^{\mu\nu} p_{l}\cdot p_{\nu_l} -i\epsilon^{\mu \alpha \nu \beta} p_{l,\alpha }p_{\nu_l,\beta} \, ,
\end{align}
where $p_l$ and $p_{\nu_l}$ are the four momenta of the lepton and the neutrino, respectively, and $E_l$ is the lepton energy in the $B$-meson rest frame.
We can simplify further by integrating the lepton energy,
\begin{align}\label{eq:inclusive diff rate}
 \frac{d^2 \Gamma}{d q^2 d q_0 } = \frac{G^2_F |V_{cb}|^2}{8\pi^3} \tilde k^{\mu\nu}(q)W_{\mu\nu} (q)\, ,
\end{align}
where for massless leptons
\begin{align}\label{eq:ktilde}
    \tilde k^{\mu\nu}(q)= \frac{|\bm q^2|q^2}{3}\left(\frac{q^\mu q^\nu}{q^2}-g^{\mu\nu}\right)\,.
\end{align}
The hadronic tensor $W_{\mu\nu}$ is defined as
\begin{align}
    W_{\mu\nu}(q)=\frac{1}{2M_B}\sum_{X_c}(2\pi)^3\delta^{(4)}(p_B-q-p_X)\langle B(\bm 0)|J_{\mu}^\dagger(0)|X_c\rangle
    \langle X_c|J_\nu(0)|B(\bm 0)\rangle\,,
\end{align}
with $J_\mu$ a $b\to c$ flavour-changing $V-A$ current. The sum is over all possible inclusive final states $X_c$ containing a charm quark. The hadronic tensor is related to the time-ordered  amplitude
\begin{align}
 H_{\mu\nu}( q) = & \, \frac{i}{2M_{B_s}} \int d^4 x \, e^{-iq\cdot x} \bra{B(\bm{0})} T[J_{\mu}^{\dagger}(x) J_{\nu}(0) ]\ket{B(\bm{0})} \,,
 \label{eq:hadronicTensor}
 \end{align}
via the optical theorem, $W_{\mu\nu}(q)=\frac 1\pi {\rm Im} H_{\mu\nu}(q)$.
Since the differential decay rate is linear in the spectral density $W_{\mu\nu}$, inclusive semileptonic decay corresponds to case I and smeared experiment and theory can be compared model-independently at finite smearing width.

\subsubsection{Smeared differential decay rate}
Experiments typically measure the doubly-differential decay rate with respect to $q_0$ and $q^2$. Within existing analysis frameworks a natural way of Poisson smearing  therefore is 
\begin{align}
    \left<\frac {d^2 \Gamma}{d q^2 d q_0}\right>_{\epsilon,q^2} 
    =&\,
    \frac{G^2_F |V_{cb}|^2}{8\pi^3} 
    \left<\tilde k^{\mu\nu}W_{\mu\nu}
    \right>_{\epsilon,q^2} \nonumber\\
    =&\,
    \frac{G^2_F |V_{cb}|^2}{8\pi^3} 
    \int\limits_{-\infty}^\infty d \omega K_\epsilon(\omega-q_0) \left[\tilde k^{\mu\nu}W_{\mu\nu}\right](\omega, \bm q^2=\omega^2-q^2)\,,\label{eq:inclusive dGAmma naive smearing}
\end{align}
i.e., smearing at constant $q^2$. This does not correspond directly to what is computed in current lattice simulations, where the smearing is carried out at constant ${\bm q}^2$ in the rest frame of the decaying hadron.
We note that the Jacobian for relating $ {d^2 \Gamma}/{ dq^2 d q_0}(q^2,q_0)$ and ${d^2 \Gamma}/{d {\bm q}^2 d q_0}(q^2,{\bm q}^2)$ is unity. One  way to reconcile both analyses therefore is to apply the modified smearing prescription
\begin{align}
    \left<\frac {d^2 \Gamma}{d q^2 d q_0}\right>_{\epsilon,\bm q^2} =&\,
    \frac{G^2_F |V_{cb}|^2}{8\pi^3} 
    \int\limits_{-\infty}^\infty d \omega K_\epsilon(\omega-q_0) \left[\tilde k^{\mu\nu}W_{\mu\nu}\right](\omega,\bm q^2)\,,
\end{align}
to experiment, i.e. the smearing at constant $\bm q^2$.
\subsubsection{Spectral reconstruction}
The lattice-QCD computation of $W_{\mu\nu}$ starts from the observation that for suitably chosen (ratios of) Euclidean correlation functions (see for instance~\cite{Barone:2023tbl}),
\begin{align}\label{eq:Laplace}
     C_{\mu\nu} (t,\bm{q})
 &=  \int\limits_{\omega_{\rm min}}^{\infty} d \omega \, W_{\mu\nu}(\omega,\bm{q}) e^{-\omega t}\,,
\end{align}
i.e., the Euclidean-time dependence is related to the hadronic tensor by means of a Laplace transform.  The lower integration limit corresponds to the energy of the lightest final state. The smeared differential decay rate is computed by first expanding
\begin{align}
    K_\epsilon(\omega-q_0)\tilde k^{\mu\nu}(q)=\sum\limits_n g^{\mu\nu}_{n}(\bm q,\epsilon)e^{-\omega na}\,,
\end{align}
where $a$ is the lattice spacing, and then observing that
\begin{align}\label{eq:inclusive reconstruction sum}
    \langle \tilde k^{\mu\nu}W_{\mu\nu}(q)\rangle_\epsilon =\sum\limits_ng_n^{\mu\nu}(\bm q,\epsilon)C_{\mu\nu}(na,\bm q)\,.
\end{align}
In this way, the smeared hadronic tensor can be reconstructed  in terms of a linear combination of Euclidean correlation functions (see details for HLT and Chebyshev approaches in~\cite{Hashimoto:2017wqo,Gambino:2020crt,Gambino:2022dvu,Barone:2023tbl,DeSantis:2025yfm,Kellermann:2025pzt}). 

Note that this proposal differs from existing implementations~~\cite{Gambino:2022dvu,Barone:2023tbl,DeSantis:2025yfm,Kellermann:2025pzt}, where the leptonic tensor with smeared-out kinematic cut-offs (Heaviside step functions implementing kinematic cuts replaced by smooth sigmoids) is used as smearing kernel provided by nature. Here we propose to leave $\tilde k^{\mu\nu}$ untouched, and implement smearing via the Poisson kernel.

\begin{figure}
    \centering
    \includegraphics[width=0.3\linewidth]{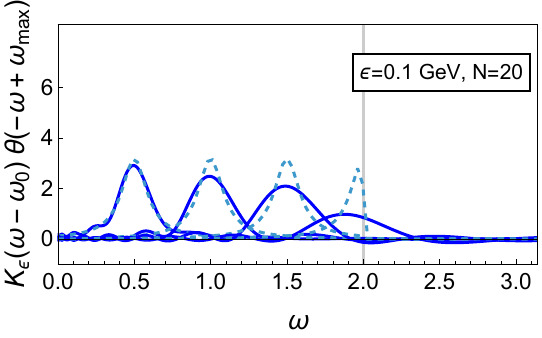}
    \includegraphics[width=0.3\linewidth]{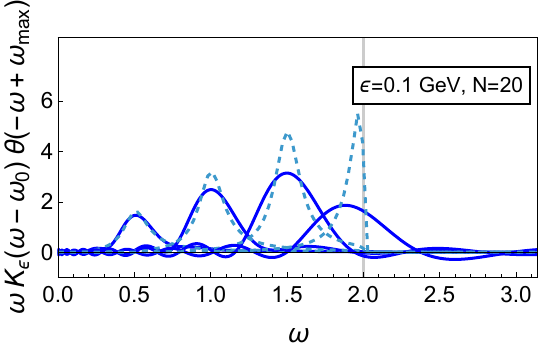}
    \includegraphics[width=0.3\linewidth]{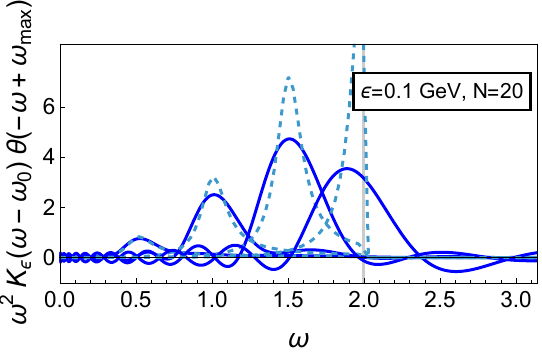}
    \includegraphics[width=0.3\linewidth]{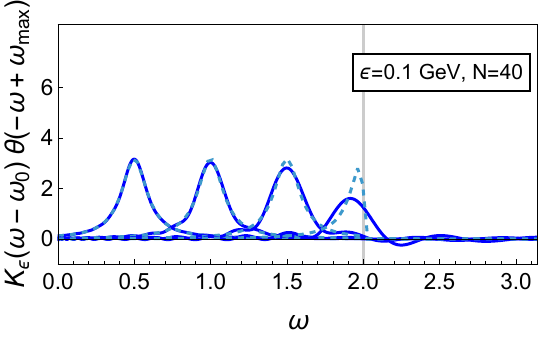}
    \includegraphics[width=0.3\linewidth]{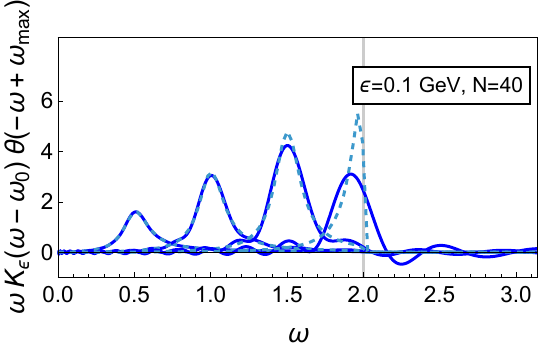}
    \includegraphics[width=0.3\linewidth]{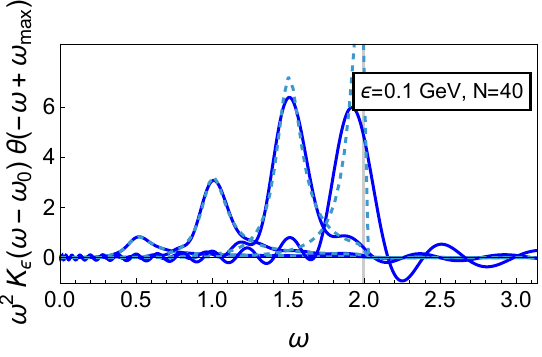}
    \includegraphics[width=0.3\linewidth]{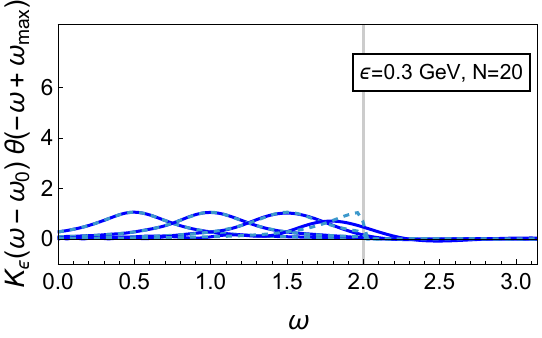}
    \includegraphics[width=0.3\linewidth]{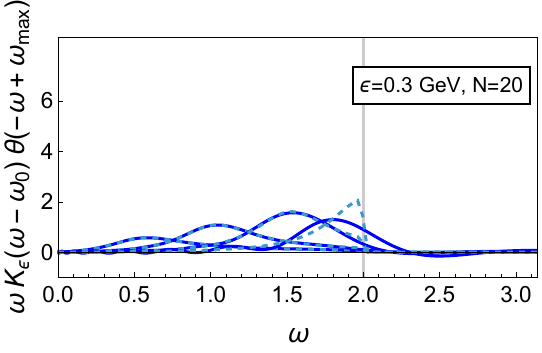}
    \includegraphics[width=0.3\linewidth]{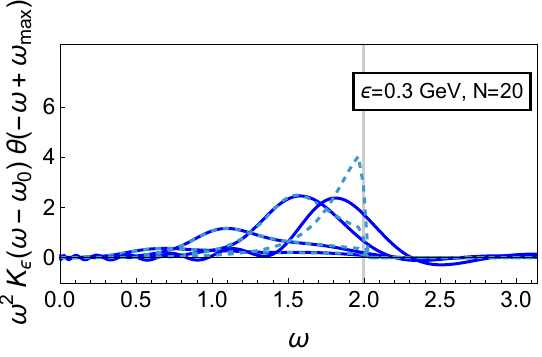}
    \includegraphics[width=0.3\linewidth]{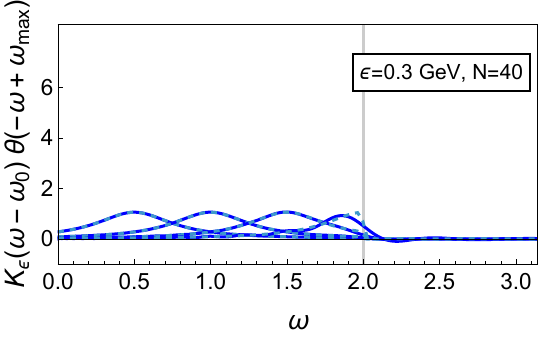}
    \includegraphics[width=0.3\linewidth]{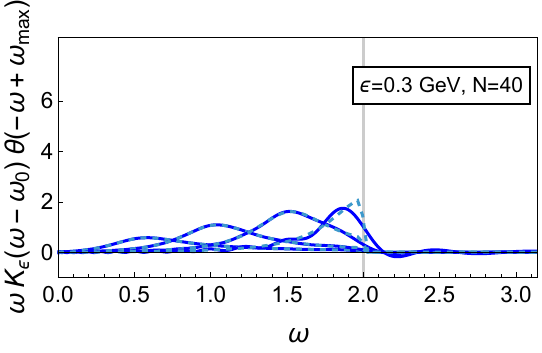}
    \includegraphics[width=0.3\linewidth]{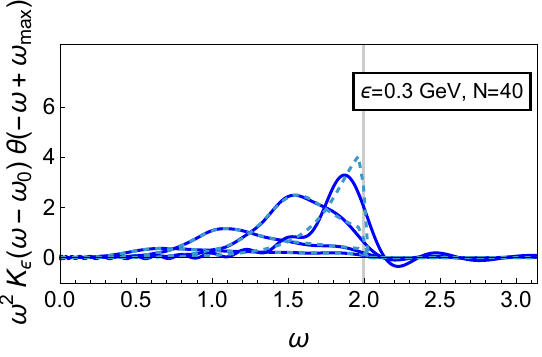}
    \caption{Plots of the kernel $\omega^n \theta(\omega_{\rm max}-\omega)K_\epsilon(\omega-q_0)
$ for $n=0,1,2$ (from left to right) with $\epsilon=100$~MeV and $\epsilon=300$~MeV with truncation 20 and 40 (see box in plot), for $q_0=0.5,1.0,1.5,2.0$~GeV and edge of the phase space at $\omega_{\rm max}=2$~GeV (vertical line). The dashed curve shows the kernel itself, while the solid lines show the approximation in terms of shifted Chebyshev polynomials that map $[\omega_{\rm min},\infty]\to [-1,1]$.}
    \label{fig:inclusiveKernel}
\end{figure}

The leptonic tensor $\tilde k^{\mu\nu}$  contains a Heaviside step function to implement the upper end of phase  space $\omega_{\rm max}=M-|\bm q|$ ($M$ is the initial-state mass and $\bm q$ the 3-momentum of the lepton-neutrino pair) and a power of energy $\omega^n$ with $n=0,1,2$. 
As an illustration for how well the kernel is described 
in terms of shifted Chebyshev polynomials of the monomial $e^{-a\omega}$ (with mapping $[\omega_{\rm min},\infty]\to [-1,1]$, see Appendix~\ref{app:Cehbyshev}),  we plot
    $\omega^n \theta(\omega_{\rm max}-\omega)K_\epsilon(\omega-q_0)
$ and the Chebyshev approximation of order 20 and 40 in the plots in Fig.~\ref{fig:inclusiveKernel} for two different smearing widths $\epsilon=100$~MeV and  $\epsilon=300$~MeV, and truncations $N=20$ and $N=40$ of the sum in Eq.~(\ref{eq:inclusive reconstruction sum}), respectively.

In particular for the wider smearing-width the approximation is excellent. If the centre $q_0$ of the kernel is chosen closer to the end of the phase space at $\omega_{\rm max}$ (vertical line in the plots), truncation effects become more pronounced. These can in principle be reduced by increasing the order of the approximation. Alternatively, tools have been developed in~\cite{Barone:2023tbl,Kellermann:2025pzt} for estimating them in a full decay-rate computation.

\subsubsection{Discussion}
A direct comparison of the Poisson-smeared differential decay rate for inclusive meson decay with correspondingly smeared lattice-QCD data is possible. A sufficiently large smearing width $\epsilon$ allows for good control of systematic effects from the continuum and  finite-volume  limits for  contemporary lattice-QCD simulations. Systematics due to the $\epsilon\to 0$ extrapolation are therefore absent. 
A computation of CKM matrix elements like $|V_{ub}|$ and  $|V_{cb}|$ for $B_{(s)}$ decay and $|V_{cs}|$ and $|V_{cd}|$ for $D_{(s)}$ decay at finite smearing width, and further SM tests based on the smeared differential decay rate or moments of it, can be accomplished. The computation at finite $\epsilon$  considerably softens the requirements on a lattice computation with controlled systematic effects.

\subsection{Example II: Rare semileptonic meson decay}
\label{sec:Semileptonic rare D decay}
While the discussion in this section concentrates on the example of the rare decay $D\to \pi \ell^+\ell^-$, the methods are applicable also to other decay channels, including $B\to K^{(\ast)} \ell^+\ell^-$. Both channels are being studied experimentally (see~\cite{LHCb:2013hxr} for the $D$ decay and~~\cite{LHCb:2013ghj,LHCb:2014cxe,LHCb:2015tgy,LHCb:2015wdu,LHCb:2015svh,LHCb:2020lmf,LHCb:2020gog,LHCb:2021xxq,LHCb:2021zwz,CMS:2024atz,LHCb:2025update} for the $B$ decay) and are highly regarded in terms of their NP reach~\cite{Bause:2020obd,deBoer:2015boa,Gisbert:2024kob,Hurth:2025vfx,Ciuchini:2022wbq,Alguero:2023jeh}. Both decays are receiving notorious long-distance contributions which are currently being modelled in the SM analysis, for instance by means of dispersive methods~\cite{Gubernari:2023puw}. In the case of $D\to\pi \ell\ell$ these effects dominate over the entire semileptonic region. A model-independent approach beyond the factorisation assumption is highly desired. This is where spectral-reconstruction techniques in lattice QCD can play a decisive role. We note that the technical aspects of a computation of the relevant long-distance effects in lattice QCD based on spectral reconstruction have recently  been discussed in~\cite{Frezzotti:2023nun,Frezzotti:2025hif}. 
The difficulties discussed there in controlling a meaningful $\epsilon\to 0$ extrapolation provide further motivation for the ideas presented here. We start by briefly summarising the relevant SM relations for semileptonic rare $D$-meson decay. The discussion carries over to rare semileptonic $B$ decays.

\subsubsection{SM expressions}
\label{sec:SM expressions rare D decay}
The differential decay rate for semileptonic rare $D$-meson decay is written as
\begin{align}
    \frac{d^2\Gamma}{dq^2\,d{\rm cos}\theta}=
    \frac {e^4} {q^4}
    \frac{\lambda^{1/2}(M_D^2,M_\pi^2,q^2)}{512\pi^3M_D^3}
    \mathcal{H}_\mu(q)L^{\mu\nu}(p_+,p_-)\mathcal{H}^{\ast}_{\nu}(q)\,,
\end{align}
where $q_\mu=(p_++p_-)_\mu=(p_D-p_\pi)_\mu$ is the momentum transfer between the initial $D$ and final  $\pi$ meson, with $p_\pm$ the charged-lepton momenta,  $\lambda(a,b,c)=a^2+b^2+c^2-2ab-2ac-2bc$, and
\begin{align}
    L^{\mu\nu}(p_+,p_-)=2\left(p_-^\mu p_+^\nu+p_-^\nu p_+^\mu-g_{\mu\nu}(p_-\cdot p_+-m_\ell^2)\right)\,,
\end{align}
the leptonic tensor for leptons of mass $m_\ell$. 
The expression for the differential decay rate simplifies after integrating out the angular dependence,
\begin{align}\label{eq:dGammadqsq}
    \frac{d\Gamma}{dq^2}=
    \frac {e^4} {q^4}
    \frac{\lambda^{1/2}(M_D^2,M_\pi^2,q^2)}{512\pi^3M_D^3}
    \mathcal{H}^\mu(q)\left[\frac {q^2-4m_\ell^2}{3}g_{\mu\nu}+\left(\frac 23 + \frac{4m_\ell^2}{3q^2}\right)q_\mu q_\nu\right] \mathcal{H}^{\ast\nu}(q)\,.
\end{align}
In the SM the amplitude $\mathcal{H}$ receives long-distance (LD) and short-distance (SD) contributions, 
\begin{align}
    \mathcal{H}_\mu(q)=H_{\rm SD,\mu}(q)+H_{\rm LD,\mu}(q)\,,
\end{align}
in the sense that $H_{\rm SD}$ can be expressed in terms of local form factors such as scalar, vector and tensor, and $H_{\rm LD}$ originates from bi-local operators of the electromagnetic current and a four-quark operator.
For convenience, we now drop the subscript LD. Since the relevant short-distance effects can be computed  using more-standard lattice techniques~\cite{Frezzotti:2025hif}, we now concentrate the discussion on long-distance contributions, returning again to the discussion of the full amplitude later on. 

The electroweak operator relevant for the long-distance contribution is $O=\lambda_d \sum\limits_{n=1,2} O^{(i)}$, with the flavour-changing four-quark operator $O^{(i)}=C_i(\mu)(O^{d,(i)}(\mu)-O^{s,(i)}(\mu))$ receiving contributions from both down and strange quarks. The current is understood  to be renormalised at renormalisation scale $\mu$, and $C_i(\mu)$ are the corresponding Wilson coefficients (cf. detailed discussion in~\cite{Frezzotti:2025hif}).
With these definitions, the relevant hadronic amplitude for the long-distance effects is
\begin{equation}\label{eq:amplitude}
    H_\mu(q)=i\int d^4x\,e^{iqx}\langle \pi(\bm p_\pi)|T[j_\mu(t,\bm x)O(0)]|D(\bm p_D)\rangle\,,
\end{equation}
where $j_\mu$ is the conserved electromagnetic current. 
The two time orderings are governed by different physics. The first time ordering (assuming $x=(t,\bm x)$) is
\begin{align}
    H_\mu^-(q)=i\int d^4x\,\theta(-t)\,e^{iqx}\,\langle \pi(\bm p_\pi)|O(0)j_\mu(t,\bm x)|D(\bm p_D)\rangle\,.
\end{align}
The charm quark originating from the $D$ meson first couples to the electromagnetic current $j_\mu$, such that the following intermediate states carry charmness $C=1$. The charm quark is converted into an up quark only after coupling to the four-quark operator $O$. The second time ordering is,
\begin{align}
    H_\mu^+(q)=&\,i\int d^4x\, \theta(t)\,e^{iqx}\langle \pi(\bm p_\pi)|j_\mu(t,\bm x)O(0)|D(\bm p_D)\rangle\,.
\end{align}
It allows for resonant intermediate states with charmness $C=0$ to appear between the four-quark operator $O$ and the electromagnetic current $j_\mu$.
Using the identity 
\begin{equation}
\theta(t)=\lim\limits_{\eta\to 0}\int\limits_{-\infty}^\infty \frac{dE}{2\pi i}\frac {e^{iEt}}{E-i\eta}\,,
\end{equation}
it can be rewritten as
\begin{align}
    H_\mu^+(q)
    =&\,i\int d^4x\, \theta(t) \langle \pi(\bm p)|j_\mu(0)
        e^{-i t(\hat H-q_0-E_\pi)}
        e^{ i\bm x\cdot (\hat{\bm P}-\bm p_D)}
        O(0)|D(\bm p_D)\rangle\nonumber\\
    =&\,\lim\limits_{\eta\to 0}\int\limits_{-\infty}^\infty\frac{dE}{2\pi}\int d^4x\,  \frac{\langle \pi(\bm p_\pi)|j_\mu(0)
        e^{-i t(\hat H-E_D-E)}
        e^{ i\bm x\cdot (\hat{\bm P}-\bm p_D)}
        O(0)|D(\bm p_D)\rangle}{E-i\eta}\nonumber\\
     =&\,\lim\limits_{\eta\to 0}\int\limits_{E_\ast^+}^\infty\frac{dE}{2\pi}\,  \frac{\rho^+_\mu(E,\bm q)}{E-E_D-i\eta}
     \qquad (E_\ast^+<E_D)\label{eq:Hmuplus}
     \,,
\end{align}
with spectral density
\begin{align}\label{eq:exclusice rho+}
    \rho^+_\mu(E,\bm q)=(2\pi)^4\langle \pi(\bm p_\pi)|j_\mu(0)
        \delta(\hat H-E)
        \delta(\hat{\bm P}-{\bm p}_D)
        O(0)|D(\bm p_D)\rangle\,.
\end{align}
Consequently, the other time ordering can be rewritten as
\begin{align}
    H_\mu^-(q)
     =&\,\lim\limits_{\eta\to 0}\int\limits_{E_\ast^-}^\infty\frac{dE}{2\pi}\,  \frac{\rho^-_\mu(E,\bm q)}{E-E_\pi-i\eta}\qquad (E_\ast^->M_D-q_0)\,,
\end{align}
with spectral density
\begin{align}\label{eq:exclusice rho-}
    \rho^-_\mu(E,\bm q)=(2\pi)^4\langle \pi(\bm p_\pi)|
        O(0)
        \delta(\hat H-E)
        \delta(\hat{\bm P}-\bm p_\pi)
        j_\mu(0)|D(\bm p_D)\rangle\,.
\end{align}
Note that we have adjusted the lower limit of the integration over $E$ to $E^{\pm}_\ast$, which corresponds to the energy of the lowest on-shell intermediate state in the respective time orderings. We observe that $E_\ast^->E_D$ and   $E_\ast^+<M_D$, respectively. The spectral density vanishes below these energies. Correspondingly, while the integration over energy in $H_\mu^-$ can be performed in the limit $\eta\to 0$, this is not the case for $H_\mu^+$, where the integration encounters poles from on-shell intermediate states. It is theses poles that hinder directly predicting the Minkowski amplitude by Wick rotation of results from Lattice-QCD simulations. 

As for the case of inclusive decay, one can now define and compute (ratios of suitable) Euclidean correlation functions 
    \begin{align}\label{eq:exclusive Laplace}
        C^+_\mu(t,\bm q)=&\,\int\limits_{E_\ast^+}^\infty {d\omega}\,e^{- (\omega-E_\pi) t}\,\rho_\mu^+(\omega,\bm q)\qquad( t>0)\,,\nonumber\\\\[-5mm]
        C^-_\mu(t,\bm q)=&\,\int\limits_{E_\ast^-}^\infty {d\omega}\,e^{+ (\omega-E_D)t}\,\rho_\mu^-(\omega,\bm q)\qquad( t\le 0)\,,\nonumber
    \end{align}
which exposes the relation to the spectral densities in terms of a Laplace transform -- the starting point for spectral reconstruction.

In~\cite{Frezzotti:2025hif} the spectral reconstruction of the amplitudes $H^\pm_\mu(\omega,\bm q)$ was discussed. Given that the energies of the initial- and final-state mesons fix the kinematics ($\omega=\omega(\bm q)$), the evaluation of Eq.~(\ref{eq:Hmuplus}) at finite smearing probes virtuality, which is not accessible experimentally. A connection to experiment can in this way only be made in the $\epsilon\to 0$ limit. Below, we therefore explore an alternative idea for the reconstruction, for which smearing of experimental data and the lattice reconstruction coincide at finite $\epsilon$.
\subsubsection{Dispersive representation}
\label{sec:Dispersive representation}
The hadronic amplitude $H_\mu(q)$ is transverse and has the Lorentz decomposition
\begin{align}
    H_\mu(\omega,\bm  q)=&\,\left(P_\mu-\frac {P\cdot q}{q^2}q_\mu\right)H_T(q^2)\equiv \,t_\mu(\omega,\bm q) H_T(s)\,,
\end{align}
where $P=p_D+p_\pi$ and $q=(\omega,\bm q)$, such that $s\equiv q^2=\omega^2-|\bm q|^2$.
This allows us to write the interference term (cf. Eq.~(\ref{eq:interference term})) as a product of two Lorentz scalars,
\begin{align}\label{eq:exclusive smeared IR}
    I^R(s)=&\,{\rm Re}\left[H_{\rm SD}^{\mu\ast}(\omega,\bm q) \phi_{\mu\nu}^\ast(\omega,\bm q)t^\nu(\omega,\bm q)H_T(s)\right]
    \equiv {\rm Re}\left[\Phi^\ast(s)H_T(s)\right]\,.
\end{align}
Throughout, we assume that the function $\Phi$ is known analytically. It contains kinematic factors, CKM matrix elements, as well as form factors of local matrix elements, which we assume to be known in terms of polynomial parametrisations~\cite{Boyd:1994tt}. It also contains a weak phase, hence, $\Phi(s)=|\Phi(s)|e^{i\phi}$.
In the same way, we can also define the imaginary part $I^I(s)$ of the interference term.
The large-$s$ behaviour of the amplitude $H_T(s)$ is such that the unsubtracted dispersion relation 
\begin{align}\label{eq:unsubtracted}
    H_T(s)=&\lim\limits_{\eta\to 0}\int\limits_{s_{\rm min}}^\infty d s'\frac{\rho_T(s')}{s'-s-i\eta}\,,
\end{align}
is not expected to converge~\cite{Feldmann:2017izn}.
Instead, one  considers the once-subtracted dispersion relation
\begin{align}\label{eq:HLD q^2 dispersion}
    H_T(s)=&\,H(q_0^2)+{(s-q_0^2)}\lim\limits_{\eta\to 0}\int\limits_{s_{\rm min}}^\infty d s'\frac{\rho_T(s')}{(s'-q_0^2)(s'-s-i\eta)}\,,
\end{align}
with subtraction point $q^2_0$, and $s_{\rm min}$ the squared energy of the lightest intermediate state. The subtraction constant $H_T(q^2_0)$ has been computed in perturbation theory with a subtraction point deep in the Euclidean region~\cite{Feldmann:2017izn,Bharucha:2020eup,Bansal:2025hcf}. 
Being a constant, it is inert to smearing. Beyond perturbation theory, one either has to find ways for computing it nonperturbatively, or, define observables for which the subtraction constant cancels. Here we discuss the latter.

Under Poisson smearing we obtain
\begin{align}\label{eq:exclusive rare Poisson vanilla}
    \langle I^R(s)\rangle_\epsilon 
    =&\,{\rm Re}\left[\Phi^\ast(s+i\epsilon)  H_T(s+i\epsilon)\right]\,,
\end{align}
and similarly for the imaginary part $I^I(s)$. 
Since, as assumed above, $\Phi$ is known analytically, it can  be evaluated off the real $s$ axis, $\Phi(s+i\epsilon)$. The local-form-factor parametrisations entering $\Phi$, such as BGL~\cite{Boyd:1994tt}, are guaranteed to converge on and off the real axis. Non-analyticities due to bound-states or the $D\pi$ production threshold $t_+=(M_D+M_\pi)^2$ are kinematically well separated from the semileptonic region, such that they can be ignored for reasonable choices of the smearing width $\epsilon< t_+-q^2_{\rm max}$.\footnote{For $D\to\pi\ell\ell$ one finds $q^2_{\rm max}\approx 3{\rm GeV}^2$ and $M_{D^\ast}^2\approx t_+\approx 4{\rm GeV}^2$, while for $B\to K\ell\ell$ one finds $q^2_{\rm max}\approx 23 {\rm GeV}^2$, $M_{B^\ast}^2\approx 28{\rm GeV}^2$ and $t_+\approx 33{\rm GeV}^2$.}
For the same reason we neglect in the following a non-vanishing phase in $\Phi$, which could originate from the admixture of the factor above $t_+$ under smearing.
%
%

%
\subsubsection{Unsubtracted dispersion relation}
While the dispersion relation for $D\to\pi \ell\ell$ decay requires one subtraction, this is not generally the case for all processes that are candidates for spectral reconstruction. For the sake of generality, and since it is the simplest case, we therefore briefly discuss how smearing acts in this case. 
The effect of smearing is
\begin{align}\label{eq:exclusive dispersive interference Re}
    \langle I^R(s)\rangle_\epsilon  =&\, 
    |\Phi^\ast(s+i\epsilon)|{\rm Re}\left[e^{-i\phi}H_T(s+i\epsilon)\right]\nonumber\\
    =&\,
   |\Phi^\ast(s+i\epsilon)|\left(\cos\phi\,{\rm Re}[H_T(s+i\epsilon)]
    +\sin\phi\,{\rm Im}[H_T(s+i\epsilon)]\right)
    \,,
\end{align}
where
\begin{align}
    {\rm Re}[H_T(s+i\epsilon)]=&\,\int\limits_{s_{\rm min}}^\infty ds'\frac{(s'-s)\rho_T(s')}{(s'-s)^2+\epsilon^2}\,,\\
    {\rm Im}[H_T(s+i\epsilon)]=&\,\int\limits_{s_{\rm min}}^\infty ds'\frac{\epsilon\rho_T(s')}{(s'-s)^2+\epsilon^2}\,.
    \end{align}
The expression in Eq.~(\ref{eq:exclusive dispersive interference Re}) can be determined from experimental data, and, as we will discuss later, also in lattice QCD.
\subsubsection{Once-subtracted dispersion relation}
In the case of the once-subtracted dispersion relation the interference term  receives an additional contribution from the subtraction constant, i.e.,
\begin{align}\label{eq:sub exclusive dispersive interference Re}
    \langle I^R(s)\rangle_\epsilon  =&\, 
  |\Phi^\ast(s+i\epsilon)|{\rm Re}\left[e^{-i\phi}\left(H_T(q^2_0=0)+F_T(s+i\epsilon)\right)\right]
    \,,
\end{align}
where
\begin{align}
    {\rm Re}F_T(s+i\epsilon)=&\,\int\limits_{s_{\rm min}}^\infty ds'\frac{(s(s'-s)-\epsilon^2)\rho_T(s')}{s'((s'-s)^2+\epsilon^2)}\,,\nonumber\\[-5mm]\label{eq:dispersive integrals}\\
    {\rm Im}F_T(s+i\epsilon)=&\,
    \epsilon \int\limits_{s_{\rm min}}^\infty ds'\frac{\rho_T(s')}{(s'-s)^2+\epsilon^2}\,,\nonumber
\end{align}
and where, for simplicity of notation but without loss of generality, we assume $q^2_0=0$.
For a meaningful comparison of smeared experiment and theory we have to devise observables for which the subtraction constant cancels. To this end, we propose to consider combinations of observables at two different kinematic points $s_1\neq s_2$
\begin{align}
    \frac{\langle I^R(s_1)\rangle_\epsilon}{{|\Phi(s_1+i\epsilon)|}}-
    \frac{\langle I^R(s_2)\rangle_\epsilon}{|\Phi(s_2+i\epsilon)|}={\rm Re}\left[e^{-i\phi}\left(F_T(s_1+i\epsilon)-F_T(s_2+i\epsilon)\right)\right]\,,
\end{align}
 or, with two different smearings $\epsilon_1\neq \epsilon_2$ 
\begin{align}
    \frac{\langle I^R(s)\rangle_{\epsilon_1}}{{|\Phi(s+i\epsilon_1)|}}-
    \frac{\langle I^R(s)\rangle_{\epsilon_2}}{|\Phi(s+i\epsilon_2)|}={\rm Re}\left[e^{-i\phi}\left(F_T(s+i\epsilon_1)-F_T(s+i\epsilon_2)\right)\right]\,.
\end{align}
These expressions are by construction independent of the subtraction constant. Next, we  discuss the reconstruction of the dispersive integrals $F_T(s+i\epsilon)$ from Euclidean correlation functions computed in lattice QCD.
\subsubsection{Kinematics for spectral reconstruction}
The objective here is to reconstruct the smeared hadronic amplitude from the  time behaviour of a Euclidean correlation function for the $D\to\pi$ transition. Since the $D$ and $\pi$ momenta are typically  fixed for a given lattice correlation function, energy conservation at finite smearing-width is broken ($\omega\neq M_D-E_\pi$, with $\omega$ the integration variable corresponding to the photon energy). In order to at least formally recover energy conservation, we introduce the additional degree of freedom $\Delta=(\omega,\bm 0)$, such that $q=p_D-p_\pi+\Delta$.\footnote{The author would like to thank Max T. Hansen for pointing out a mistake in arguments related to this point in an earlier version of this manuscript.} 
In the limit of vanishing smearing width, $\Delta$ vanishes and energy-conservation is recovered. 
At finite smearing, the additional degree of freedom requires introducing a further invariant in the Lorentz decomposition of the hadronic amplitude (see App.~\ref{app:transversality off-shell} for more details),
\begin{align}
    H_\mu(\omega,\bm p)=F_1(q^2,\Delta^2)T_{1,\mu}(\omega,\bm p)+
    F_2(q^2,\Delta^2)T_{2,\mu}(\omega,\bm p)\,,
\end{align}
where
\begin{align}
    T_{1,\mu}(\omega,\bm p)=P_\mu-\frac {P\cdot q}{q^2}q_\mu\,,\qquad{\rm and}\qquad
    T_{2,\mu}(\omega,\bm p)=\Delta_\mu-\frac {\Delta\cdot q}{q^2}q_\mu\,.
\end{align}
As the smearing is removed, $T_2$ vanishes, and we identify $H_T(q^2)=F_1(q^2,0)$ as the physical amplitude. We therefore introduce the vector
\begin{align}\label{eq:tbar definition}
    \bar t_\mu(\omega,\bm p)=\mathcal{N} \left(T_{1,\mu}-\frac{T_1\cdot T_2}{T_2\cdot T_2}T_{2,\mu}\right)\,,
\end{align}
with
\begin{align}
    \mathcal{N}^{-1}=T_1\cdot T_1-\frac{(T_1\cdot T_2)^2}{T_2\cdot T_2}\,,
\end{align}
such that 
\begin{align}\label{eq:projected amplitude}
    F_1(q^2,\Delta^2)=\bar t_\mu(\omega,\bm p)H^\mu(\omega,\bm p)\,.
\end{align}
projects on the physical form factor.
In the next section we propose a combined energy-momentum smearing, in order to control the effect of a non-vanishing $\Delta$. 
\subsubsection{Energy-momentum spectral reconstruction}
We  continue discussing the subtracted dispersion relation, which ensures convergence of the dispersive integral for rare semileptonic meson decay. In the $D$-meson rest frame, energy conservation imposes 
$p_\omega^2\equiv|\bm p_\pi|^2=(M_D-\omega)^2-M_\pi^2$ for
the pion momentum.
We can therefore rewrite the integration over squared momentum transfer $s'=\omega^2-p_\omega^2$ in Eq.~(\ref{eq:dispersive integrals}) into one over  $\omega$ as
\begin{align}\label{}
    {\rm Re}
F_T(s+i\epsilon)=&\,2M_D\int\limits_{{\omega_{\rm min}}}^\infty d\omega\,\frac{(s(\omega^2-p_\omega^2-s)-\epsilon^2)\rho_T(\omega^2-p_\omega^2)}
    {(\omega^2-p_\omega^2)((\omega^2-p_\omega^2-s)^2+\epsilon^2)}\,,\nonumber\\
        {\rm Im}F_T(s+i\epsilon)=&\,\epsilon2M_D
    \int\limits_{\omega_{\rm min}}^\infty d\omega\frac{ \rho_T(\omega^2-p_\omega^2)}
    {(\omega^2-p_\omega^2-s)^2+\epsilon^2}\nonumber\,.
\end{align}
We will in the following use the more compact notation
\begin{align}\label{}
    {\rm Re}F_T(s+i\epsilon)\equiv&
    \int\limits_{\omega_{\rm min}}^\infty d\omega\,f^R_{s,\epsilon}(\omega,p_\omega)=
   \int\limits_{\omega_{\rm min}}^\infty d\omega
    \int 
    {dp}\,f^R_{s,\epsilon}(\omega,p)\delta(p-p_\omega)\,,
\end{align}
and correspondingly for ${\rm Im}F_T$ with $f^I_{s,\epsilon}(\omega,p_\omega)$, where, for reasons to become clear next, we have added an additional \emph{trivial} integration over momentum $p$, which imposes the dispersion relation.

\begin{figure}[t!]
\begin{center}
 \includegraphics[width=7cm]{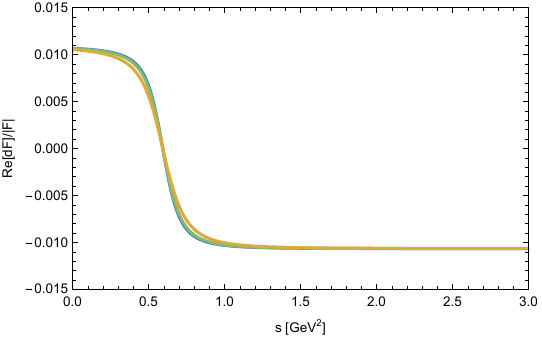}
 \includegraphics[width=7cm]{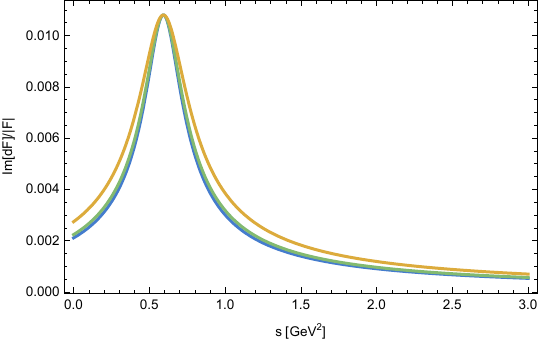}\\[4mm]
 \includegraphics[width=7cm]{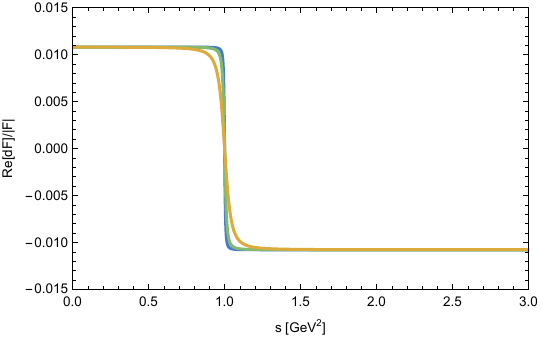}
 \includegraphics[width=7cm]{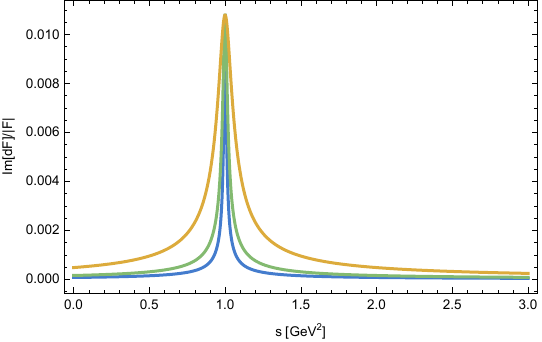}
 \end{center}
 \caption{Relative error for real (left) and imaginary (right) parts of the amplitude assuming $\sigma\approx 0.2$GeV for the $\rho$ (top) and $\phi$ (bottom) meson assuming three different values for the energy smearing $\epsilon=(\{0.05,0.1,0.5\}{\rm GeV})^2$ (blue, green, yellow).
 }
 \label{fig:dF}
\end{figure}

We now allow for small deviations from the correct dispersion relation $p_\omega$ by replacing the Dirac-delta by a Gaussian damping term:
\begin{align}\label{eq:Idef}
    {\rm Re}F_{T,\sigma}(s+i\epsilon)=\int\limits_{\omega_{\rm min}}^\infty d\omega \int dp\, f^{R}_{s,\epsilon}(\omega, p)R_\sigma(p-p_\omega)\,,
\end{align}
(and similarly for the imaginary part) where  
\begin{align}
    R_\sigma(p-p_\omega)=\frac 1{\sqrt{2\pi\sigma^2}}e^{-(p-p_\omega)^2/(2\sigma^2)}\,,
\end{align}
with the  width $\sigma=\mathcal{O}(2\pi/L)$, and $L$ is the spatial extend of the lattice. 
We  estimate the induced systematic in terms of the 
correction term
\begin{align}\label{eq:dF}
    dF^{R/I}_{\epsilon,\sigma}(s+i\epsilon)=\int\limits_{\omega_{\rm min}}^\infty d\omega \int d\eta\,f^{R/I}_{s,\epsilon}(\omega,p_\omega+\eta)\left[R_\sigma(\eta)-\delta(\eta)\right]\,,
\end{align}
where $\eta=p-p_\omega$. The integral will be dominated by values of $\eta$ within a few multiples of $\sigma$ around zero. We therefore expand 
\begin{align}
    f^{R/I}_{s,\epsilon}(\omega,p_\omega+\eta)
    =f^{R/I}_{s,\epsilon}(\omega,p_\omega)+
    (f^{R/I})'_{s,\epsilon}(\omega,p_\omega)\eta
    +\frac 12 (f^{R/I})''_{s,\epsilon}(\omega,p_\omega)\eta^2+\dots\,.
\end{align}
The contribution to the momentum integral Eq.~(\ref{eq:dF}) from the leading term cancels between $R_\sigma$ and the Dirac-$\delta$ term, and the $(f^{R/I})'$ term, which is linear in $\eta$, vanishes due to the symmetry of $R_\sigma$. 
We therefore expect the correction to be
 \begin{align}
     dF^{R/I}_{\epsilon,\sigma}(s+i\epsilon)=\frac{\sigma^2}{2}\int d\omega (f^{R/I})''_{s,\epsilon}(\omega,p_\omega)+\mathcal{O}(\sigma^4)\,,
 \end{align}
 i.e., the correction predominantly depends on the curvature of the dispersive kernel at the evaluation point $s$, suppressed by two powers of the  momentum-smearing width $\sigma$. We note that:
 \begin{itemize}
     \item $(f^{R/I})''_{s,\epsilon}$ is the curvature of the smeared  dispersive kernel following the correct dispersion relation. In order to estimate corrections due to the finite momentum-smearing width $\sigma$, we can therefore employ models of the physical spectral density.
     \item For narrow resonances like $\phi$, where in practice the smearing radius will be larger than the resonance width, $\epsilon\gg \Gamma$, the smearing will smoothen large curvature and therefore be the determining factor for the magnitude of the correction term. 
 \end{itemize} 
The proposal allows for deviations from physical kinematics governed by the momentum-smearing width $\sigma$. For an estimate of the size of the systematic effects introduced in this way, we model the dispersive kernel in Eq.~(\ref{eq:Idef}) in terms of a Breit-Wigner, i.e.,
\begin{align}
f_{s',\epsilon}(\omega,p_\omega)=\frac{1}{\omega^2-p_\omega^2-s'-i\epsilon}\frac 1\pi\frac{M\,\Gamma}{(M-(\omega^2-p_\omega^2))^2+M^2\Gamma^2}\,.
\end{align}
We consider the two examples of the $\rho$ and $\phi$ resonances with $M_\rho=0.77$GeV, $\Gamma_\rho=0.15$GeV and $M_\phi= 1.019$GeV, $\Gamma_\phi=0.004$GeV, respectively. We exemplarily use the smearing parameters $\epsilon=(\{0.05,0.1,0.5\}{\rm GeV})^2$. For charm quarks to be well controlled in lattice simulations requires a lattice cutoff of, say, $a^{-1}=3$GeV. Keeping at the same time exponentially suppressed finite-volume effects at bay requires $M_\pi L\approx 4$. With the pion mass in units of the lattice spacing of $aM_\pi\approx 0.045$, this corresponds to $L/a\approx 89$. A typical nearby simulation parameter is $L/a=96$. A first choice for $\sigma$ could therefore be $\sigma=2\pi/L\approx 0.2{\rm GeV}$. 
In Figure~\ref{fig:dF} we plot the real and imaginary parts of the relative error
$dF_{\epsilon,\sigma}(s)/|F_{\epsilon,\sigma}(s)|$, respectively. The plots indicate that the errors are sufficiently small for making meaningful SM predictions in the presence of both $\rho$ and $\phi$ resonances.

With the energy-momentum-smearing strategy now introduced, we can now reconstruct the interference term using the smearing kernels for the real and imaginary part
\begin{align}
    \mathcal{K}^R_{\epsilon,\sigma,\mu}(s,\omega,\bm p)=&
    \, 
   \theta(q^2_{\rm max}-q^2_{\omega,\bm p})
   R_\sigma(|\bm p|^2-p_\omega)
   2M_D\,\bar t_\mu(\omega,\bm p)\,
   \frac{s(q^2_{\omega,\bm p}-s)-\epsilon^2}{q^2_{\omega,\bm p}((q^2_{\omega,\bm p}-s)^2+\epsilon^2)}
    \,,\label{eq:excluive kernel re}\\
    \mathcal{K}^I_{\epsilon,\sigma,\mu}(s,\omega,\bm q)=&\,
    \,
    \theta(q^2_{\rm max}-q^2_{\omega,\bm p})
    R_\sigma(|\bm p|^2-p_\omega)
    \epsilon2M_D\,\bar t_\mu(\omega,\bm p)\,
    \frac{ \,1}{(q^2_{\omega,\bm p}-s)^2+\epsilon^2}
     \,,\label{eq:excluive kernel im}
\end{align}
where $q^2_{\omega,\bm p}=\omega^2-|\bm p|^2$, which are modelled after the dispersive integrals in Eq.~(\ref{eq:dispersive integrals}). The kernels are centered around the kinematic points $s$ and $p_\omega$, respectively, and smear in energy $\omega$ and momentum $\bm p$. 
The kernel further makes use of the 
 transversality of the photon amplitude, which following
in Eq.~(\ref{eq:projected amplitude}) allows us to
write $\rho_T(s)=\bar t^\mu(\omega,\bm p)(\rho_\mu^+(\omega,\bm p)+\rho_\mu^-(\omega,\bm p))$. Note that individually, $\rho^+_\mu$ and $\rho^-_\mu$ are not transversal. Only their sum is.
The Heaviside step function with $q^2_{\rm max}=(M_D-M_\pi)^2$ restricts the smearing to the semileptonic region. 
The kernels can be expanded as\footnote{In practice, it is advisable to replace the Heaviside in Eqs.~(\ref{eq:excluive kernel re}) and (\ref{eq:excluive kernel im}) by a sigmoid with a width $\delta\ll\epsilon$.}
\begin{align}\label{eq:exclusive reconstruction kernels}
   \mathcal{K}^{R/I}_{\epsilon,\sigma,\mu}(s,\omega,\bm q)=
   \sum\limits_{t=t_\pm}\sum\limits_{\bm x} c^{R/I,\pm}_{t,\bm x,\mu}(s,\epsilon,\sigma)e^{-\omega t}e^{i\bm p\cdot \bm x}\,,
\end{align}
with $t_+=1$ and $t_-=0$.
The reconstruction of $F(s+i\epsilon)$  is done simultaneously with data from both time orderings. We then obtain
\begin{align}\label{eq:reconstructed I}
\langle {\rm Re}[\Phi^\ast(s) H_T(s)]\rangle_\epsilon=
|\Phi^\ast(s+i\epsilon)|e^{i\phi}\nonumber\\
\times\Bigg\{H_T(0)
    +\sum_{X=\pm}\sum_{t=t_X}\sum_{\bm p}\Big(
    &\, c^{R,X}_{t,\bm p,\mu}(s,\epsilon,\sigma)
    \,e^{E_Xt}+
    i\, c^{I,X}_{t,\bm p,\mu}(s,\epsilon,\sigma)
    \,e^{E_Xt}\Big)\,C_\mu(\pm t,\bm p)\Bigg\}\,,
\end{align}
where $E_+=-E_\pi$ and $E_-=+E_D$. Note that the expression still contains the unknown subtraction constant $H_T(0)$. The sum over $\bf x$ in Eq.~(\ref{eq:exclusive reconstruction kernels})  has been replaced by a sum over pion momenta $\bf p$ using the orthogonality of plane waves $e^{i{\bf p}\cdot {\bf x}}$. Following the proposal in the previous section, the evaluation of Eq.~(\ref{eq:reconstructed I}) at two different values for $s$, or, with two different smearing widths allows to eliminate the constant. 
 The simultaneous reconstruction from both time orderings guarantees the cancellation of contact terms, and hence, the
 transversality of the spectral function. The remaining renormalisation program follows~\cite{Frezzotti:2025hif}.
\begin{figure}
    \centering
    \includegraphics[width=0.45\linewidth]{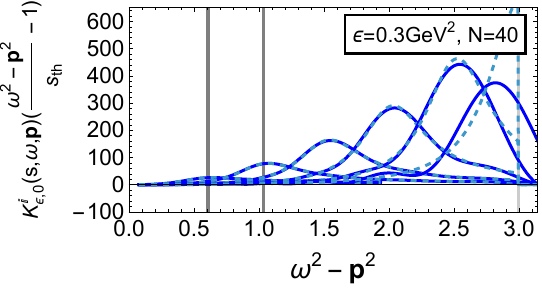}
    \includegraphics[width=0.45\linewidth]{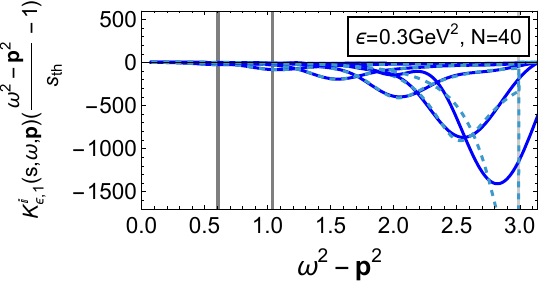}\\
    \includegraphics[width=0.45\linewidth]{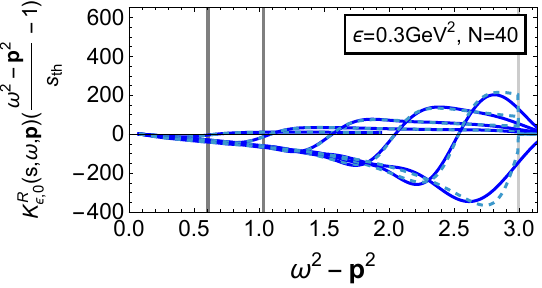}
    \includegraphics[width=0.45\linewidth]{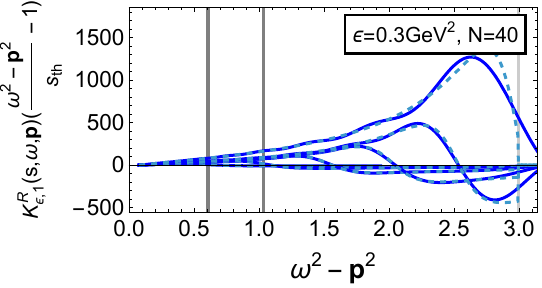}
    \caption{Plots of the real (top) and imaginary (bottom) part of the kernel as defined in Eq.~(\ref{eq:excluive kernel re}) and (\ref{eq:excluive kernel im}) for $\epsilon=0.3{\rm GeV}^2$. The plots on the left (right) show the kernel for the time (a spatial) component of the projector $\bar t_\mu$, rescaled with the expected threshold behaviour of the $D\to\pi \ell\ell$ spectral density.
    The dashed lines show the kernel and the solid lines are Chebyshev approximations (mapping energies $[\sqrt{s_{\rm th}},\infty]\to [-1,1]$) of order $N=40$. We show the cases $s=0.5,1.0,1.5,2.0,2.5,3.0\,{\rm GeV}^2$. Vertical lines indicate (from left to right) 
    $M_\rho^2$, $M_\omega^2$, $M_\phi^2$ and $q^2_{\rm max}=(M_D-M_\pi)^2$. Note that $s_{\rm th}$ can be chosen freely between 0 and the value corresponding to the energy of the lowest threshold (here $4M_\pi^2$).}
    \label{fig:RareExclusiveKernel}
\end{figure}
Fig.~\ref{fig:RareExclusiveKernel} shows the kernels in Eq.~(\ref{eq:exclusive reconstruction kernels}) for $\sigma=0$ together with the approximation in terms of Chebyshev polynomials (see Appendix~\ref{app:Cehbyshev}). For the exemplary smearing width of $\epsilon=0.3{\rm GeV}^2$ the polynomial of order 40 provides an excellent approximation to the smearing kernel. As for the inclusive decay, the truncation errors can be studied systematically within the  Chebyshev-reconstruction~\cite{Kellermann:2025pzt}. As the centre $s$ of the smearing is chosen closer to $q^2_{\rm max}=(M_D-M_\pi)^2$, the reconstructed kernel is less capable of reproducing the kinematical cut at $q^2_{\rm max}$ and \emph{leaks} beyond the kinematically allowed (for semileptonic decay) region. This leakage leads to corrections to Eq.~(\ref{eq:exclusive rare Poisson vanilla}), which can be reduced by increasing the order of the expansion. There is however little leakage when the smearing is centred in the interesting region of the $\phi$, $\rho$ and $\omega$ resonances, for which $q^2_{\rm max}-s\gg\epsilon$, and indeed, $t_+-s\gg\epsilon$.

\subsubsection{Discussion}
We have shown that the Poisson-smeared pure short-distance and short-long-distance interference contributions to rare semileptonic meson decay can be directly compared between experiment and theory. To this end we propose a smearing prescription that by exploiting the transversality of the long-distance photon amplitude allows for smearing both lattice and experiment along the same kinematic trajectory. Since such a comparison is not possible for pure long-distance contributions $\sim H_{\rm LD}H_{\rm LD}^\ast$ (unless model assumptions are made), the analysis has to concentrate on experimental observables,  such as CP asymmetries, where these contributions are absent. Since systematic effects in lattice-QCD computations at finite smearing width $\epsilon$ are under much better control, model-independent SM analyses that correctly take into account resonance effects are now within reach. 
\section{Conclusions}
\label{sec:Conclusions}
The development of spectral-reconstruction methods has opened new opportunities for lattice-QCD calculations of phenomenologically relevant quantities. For instance, the hadronic amplitudes describing inclusive meson decay and long-distance contributions to rare semileptonic meson decay, and many other processes, are now coming within reach of lattice simulations. 

Since lattice calculations in a finite  volume can only reconstruct energy-smeared spectral densities, the smearing width sets a new infrared scale. It needs to be large enough to cover the separations between neighbouring peaks of the discrete finite-volume spectral density. At the same time, it has to be smaller than the typical interval in the energy across which the amplitude varies significantly. Only if the lattice simulations entering the  continuum and infinite-volume limit fulfil these conditions and the limits have been taken, a controlled extrapolation to vanishing smearing width is possible, where the observable as measured in experiment is recovered.

Since these requirements can be rather challenging to meet  for phenomenologically relevant processes, this paper proposes to carry out SM tests completely model-independently at finite smearing width. In this approach the costly and potentially model-dependent extrapolation of lattice results to vanishing smearing width is avoided. In fact, also the control of other lattice systematics, in particular finite-volume errors is facilitated. We studied smearing with a Poisson kernel and found that model-independent comparisons are possible for observables that are linear in the spectral density (case I, e.g. inclusive meson decay). For these cases we showed that the smeared experimental observable corresponds to the harmonic extension of the amplitude, which in turn corresponds to a lattice-reconstructed observable at the same finite smearing width. To this end, we make a concrete proposal for a smearing prescription and the reconstruction for the case of inclusive semileptonic meson decay, which differs from what is currently being used~\cite{Hashimoto:2017wqo,Gambino:2020crt,Gambino:2022dvu,Barone:2023tbl,DeSantis:2025yfm,Kellermann:2025pzt}. 

The comparison of pure long-distance contributions to smeared observables that are bilinear in the hadronic amplitude is less straight-forward (case II, e.g. rare semileptonic decay), due to irreducible corrections that can be particularly strong in the vicinity of resonances. Using Breit-Wigner as a model for resonance contributions allowed us to gain analytical understanding of the  corrections (defect term). Based on the properties of the defect (positive-definiteness) and properties of the smearing kernel (semi group), we devised novel ways to test the SM based on smeared experiment and theory even in the presence of pure long-distance contributions. 

The complications arising from the pure long-distance contribution can be avoided altogether for observables where they cancel, such as CP asymmetries. The long-distance contribution then appears only linearly, namely in interference terms. We propose a new reconstruction prescription that directly mimics the effect of smearing of experimental data by exploiting the transversality of the photon amplitude, thereby allowing for finite-smearing-width comparison of theory with experiment. This new strategy avoids the computationally and technically very demanding limit of vanishing smearing width of prior proposals, opening the door to a powerful new class of SM tests.

The ideas discussed here are, in principle, applicable to a wider set of observables and processes, such as scattering of hadrons, or hadronic decay -- more generally, quantities for which intermediate on-shell hadrons contribute, and where experiment and theory can access kinematic parameters that allow for devising a suitable smearing prescription.
It's important to note that  progress in this direction will
 require close collaboration between experimentalists and theorists.\\
  
\noindent{\bf Acknowledgments:} 
The author would like to thank, Matteo Di Carlo, Jonathan Flynn, Max T. Hansen, Shoji Hashimoto, Gudrun Hiller,  Dominik Mitzel, Markus Prim and Chris Sachrajda for really helpful and illuminating discussions. 
The author is
supported by the Eric \& Wendy Schmidt Fund for Strategic Innovation (grant agreement SIF-2023-004).

 \begin{appendix}
 \section{Shifted Chebyshev polynomials}\label{app:Cehbyshev}
Shifted Chebyshev polynomials $\tilde{T}_k(x)$  are defined on a semi-infinite interval 
$[\omega_0, \infty)$ via the composition \begin{equation}
\tilde{T}_k(x) = T_k(h(x))\,,\end{equation}
where 
$h(x) = Ae^{-x} + B$ is an exponential map onto $[-1,1]$ with coefficients 
$A = -2e^{\omega_0}$ and $B=1$ chosen to satisfy $h(\omega_0) = -1$ and $h(\infty) = +1$. 
They inherit the minimax optimality and boundedness $|\tilde{T}_k(x)| \leq 1$ of the 
standard Chebyshev polynomials, and admit a convenient representation as finite sums 
\begin{equation}
    \tilde{T}_k(x) = \sum_{j=0}^k \tilde{t}^{(k)}_j e^{-jx}\,,
\end{equation}making them well-suited for 
polynomial approximation of functions defined in terms of $e^{-x}$ on an unbounded domain.
See further discussion in appendix A.2 of~\cite{Barone:2023tbl}.
\section{Transversality off-shell} \label{app:transversality off-shell}
In the absence of smearing, we can use the four vector 
\begin{align}
    t_\mu=P_\mu-\frac {P\cdot q}{q^2}q_\mu\,,
\end{align}
to relate the  amplitude $H^\mu$ to its transverse pendant, $H_T(q^2)\sim t_\mu(M_D-E_\pi(\bm p),\bm p)H^\mu(M_D-E_\pi(\bm p),\bm p)$, where we assume the $D$-meson rest frame with $\bm p$ the pion momentum. Thanks to charge conservation, $H_T$ is the only invariant form factor.

When reconstructing from a Euclidean 4pt function on the lattice, the meson three-momenta, and therefore their energies, are fixed. Under the dispersive integral over energies $\omega$ at finite smearing width, energy conservation is therefore violated.
In order to at least formally recover energy conservation, we introduce the additional degree of freedom $\Delta=(\omega,\bm 0)$, such that $q=p_D-p_\pi+\Delta$. 
In the limit of vanishing smearing width the vector $\Delta$ vanishes and energy-conservation is recovered. 
At finite smearing, the following invariants then exist:
\begin{align} p_\pi^2=M_\pi^2,\,p_D^2=M_D^2,\,q^2,\,\Delta^2,\,p_D\cdot q,\,p_\pi\cdot q,\,p_D\cdot p_\pi\,.
\end{align}
One  invariant can be eliminated via
\begin{align}
    \Delta^2=q^2+M_D^2+M_\pi^2+2\left(-p_D\cdot q+p_\pi\cdot q-p_D\cdot p_\pi\right)\,.
\end{align}
With $p_D\cdot p_\pi$ being fully determined by the fixed pion and kaon three-momenta, 
we are therefore left with the  invariants
\begin{align}   
q^2,\,\Delta^2,\,p_D\cdot q\,.
\end{align}
In the $D$-meson rest frame, where $p_D\cdot q=M_D^2-p_D\cdot p_\pi+M_D\omega$, the kinematic dependence of the form factors is then fully    described by $q^2$ and $\Delta^2=\omega^2$.
The Lorentz decomposition of the hadronic amplitude is then
\begin{align}
    H_\mu(\omega,\bm p)=F_1(q^2,\Delta^2)p_{\pi,\mu}+
    F_2(q^2,\Delta^2)p_{D,\mu}+
    F_3(q^2,\Delta^2)\Delta_\mu\,.
\end{align}
Current conservation ($q_\mu H^\mu(\omega,\bm p)=0$), reduces the number of independent invariant form factors to two,
\begin{align}
    H_\mu(\omega,\bm p)=\bar F_1(q^2,\Delta^2)t_{\pi,\mu}+
    \bar F_2(q^2,\Delta^2)t_{D,\mu}\,,
\end{align}
in this basis expressed in terms of two of the three transverse structures
\begin{align}
    t_{\pi,\mu}=&\,p_{\pi,\mu}-\frac{p_\pi\cdot q}{q^2}q_\mu\,,\nonumber\\
    t_{D,\mu}=&\,p_{D,\mu}-\frac{p_D\cdot q}{q^2}q_\mu\,,\\
    t_{\Delta,\mu}=&\,\Delta_{\mu}-\frac{\Delta\cdot q}{q^2}q_\mu\,,
\end{align}
for which
\begin{align}
    t_{D,\mu}-t_{\pi,\mu}+t_{\Delta,\mu}=0\,.
\end{align}
The three structures are, hence, not independent. Identifying
\begin{align}
    T_{1,\mu}(\omega,\bm p)=&\,P_\mu-\frac {P\cdot q}{q^2}q_\mu=t_{\pi,\mu}+t_{D,\mu}=2t_{D,\mu}+t_{\Delta,\mu}\\
    T_{2,\mu}(\omega,\bm p)=&\,\Delta_\mu-\frac {\Delta\cdot q}{q^2}q_\mu=t_{\pi,\mu}-t_{D,\mu}=t_{\Delta,\mu}\,,
\end{align}
we can then write
\begin{align}
    H_\mu(\omega,\bm p)=&\,\frac 12 \bar F_1(q^2,\Delta^2)\left(T_{1,\mu}(\omega,\bm p)+T_{2,\mu}(\omega,\bm p)\right)+
    \frac 12 \bar F_2(q^2,\Delta^2)
    \left(T_{1,\mu}(\omega,\bm p)-T_{2,\mu}(\omega,\bm p)\right)\nonumber\\
    =&\, \tilde F_1(q^2,\Delta^2)T_{1,\mu}(\omega,\bm p)+
    \tilde F_2(q^2,\Delta^2)
    T_{2,\mu}(\omega,\bm p)\,,
\end{align}
where $\tilde F_1=\frac 12 (\bar F_1+\bar F_2)$ and $\tilde F_2=\frac 12(\bar F_1-\bar F_2)$, respectively.
As the smearing is removed, the projector $T_2\sim t_{\Delta}$ vanishes, and we identify $H_T(q^2)=\tilde F_1(q^2,0)$ as the physical amplitude. We therefore propose to use the vector
\begin{align}
    \bar t_\mu(\omega,\bm p)=\mathcal{N} \left(T_{1,\mu}-\frac{T_1\cdot T_2}{T_2\cdot T_2}T_{2,\mu}\right)\,,
\end{align}
with
\begin{align}
    \mathcal{N}^{-1}=T_1\cdot T_1-\frac{(T_1\cdot T_2)^2}{T_2\cdot T_2}\,,
\end{align}
during the reconstruction, such that 
\begin{align}
    \tilde F_1(q^2,\Delta^2)=\bar t_\mu(\omega,\bm p)H^\mu(\omega,\bm p)\,.
\end{align}
\end{appendix}

\bibliographystyle{JHEP}
\bibliography{biblio}
\end{document}